\documentclass[runningheads]{llncs}
\usepackage{multirow}
\usepackage{hyperref}
\usepackage{subcaption}
\usepackage{graphicx}
\begin{document}
\title{The Memento Tracer Framework:\\Balancing Quality and Scalability for Web Archiving}

\titlerunning{The Memento Tracer Framework}
%
\author{
Martin Klein\inst{1}\orcidID{0000-0003-0130-2097} \and
Harihar Shankar\inst{1}\orcidID{0000-0003-4949-0728} \and
Lyudmila Balakireva\inst{1}\orcidID{0000-0002-3919-3634} \and
Herbert Van de Sompel\inst{2}\orcidID{0000-0002-0715-6126}}
\authorrunning{M. Klein et al.}
\institute{Los Alamos National Laboratory, Los Alamos, NM 87545, USA \\
\email{\{mklein,harihar,ludab\}@lanl.gov} \and
Data Archiving and Networked Services, Anna van Saksenlaan 51, 2593 HW Den Haag, The Netherlands \\
\email{herbert.van.de.sompel@dans.knaw.nl}}
\maketitle              
\begin{abstract}
Web archiving frameworks are commonly assessed by the quality of their archival records and by their ability to
operate at scale. The ubiquity of dynamic web content poses a significant challenge for crawler-based solutions
such as the Internet Archive that are optimized for scale. Human-driven services such as the Webrecorder tool
provide high-quality archival captures but are not optimized to operate at scale. We introduce the Memento Tracer
framework that aims to balance archival quality and scalability. We outline its concept and architecture and
evaluate its archival quality and operation at scale. Our findings indicate quality is on par or better 
compared against established archiving frameworks and operation at scale comes with a manageable overhead.
\end{abstract}
\section{Introduction and Motivation}
The web archiving landscape has evolved significantly over the last twenty years. While the Internet Archive (IA) is the uncontested 
pioneer in this field and is to date by far the largest publicly available web archive, we are now able to freely access archived web resources 
from more than twenty web archives around the world\footnote{\url{http://timetravel.mementoweb.org/}}. Many national libraries and archives 
such as the National Library of Australia \cite{nla} and the UK National Archives \cite{ukna} have begun to capture parts of the web and 
contribute to increased diversity.
However, most of the current web archiving frameworks are optimized either to try to cope with the scale of the web or to provide high-quality 
archival captures. The IA, for example, generally crawls the web in a best-effort approach, capturing as many web resources as possible. 
This results in an ever-increasing number of URIs archived and available via the IA's Wayback Machine replay engine \cite{wayback}. At the time 
of writing, this number stands at more than $731$ billion URIs \cite{kahle:tweet}. 

\begin{figure}[t!]
 \centering
 \begin{subfigure}[h]{0.49\textwidth}
  \includegraphics[scale=0.15]{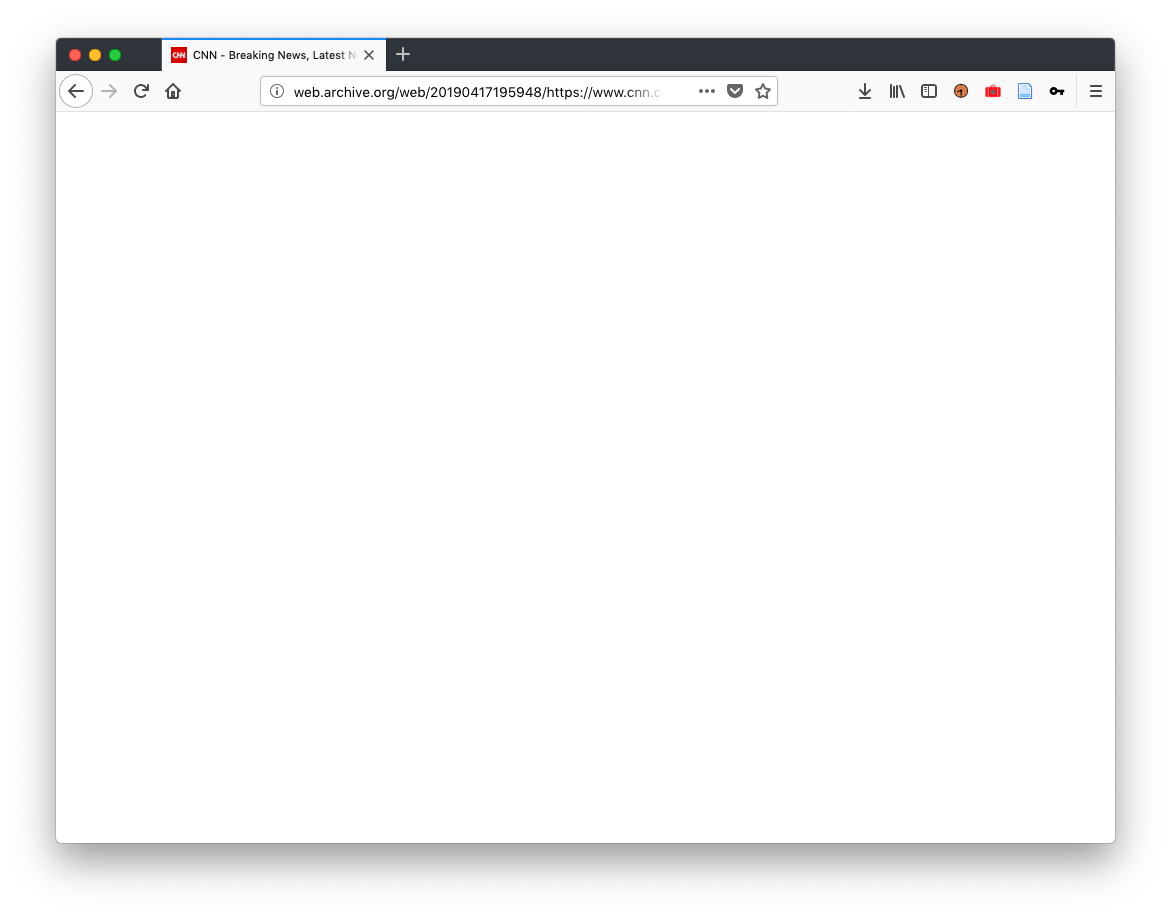}
  \caption{Internet Archive}
  \label{fig:cnn_ia_replay}
 \end{subfigure}
 \begin{subfigure}[h]{0.49\textwidth}
 \includegraphics[scale=0.15]{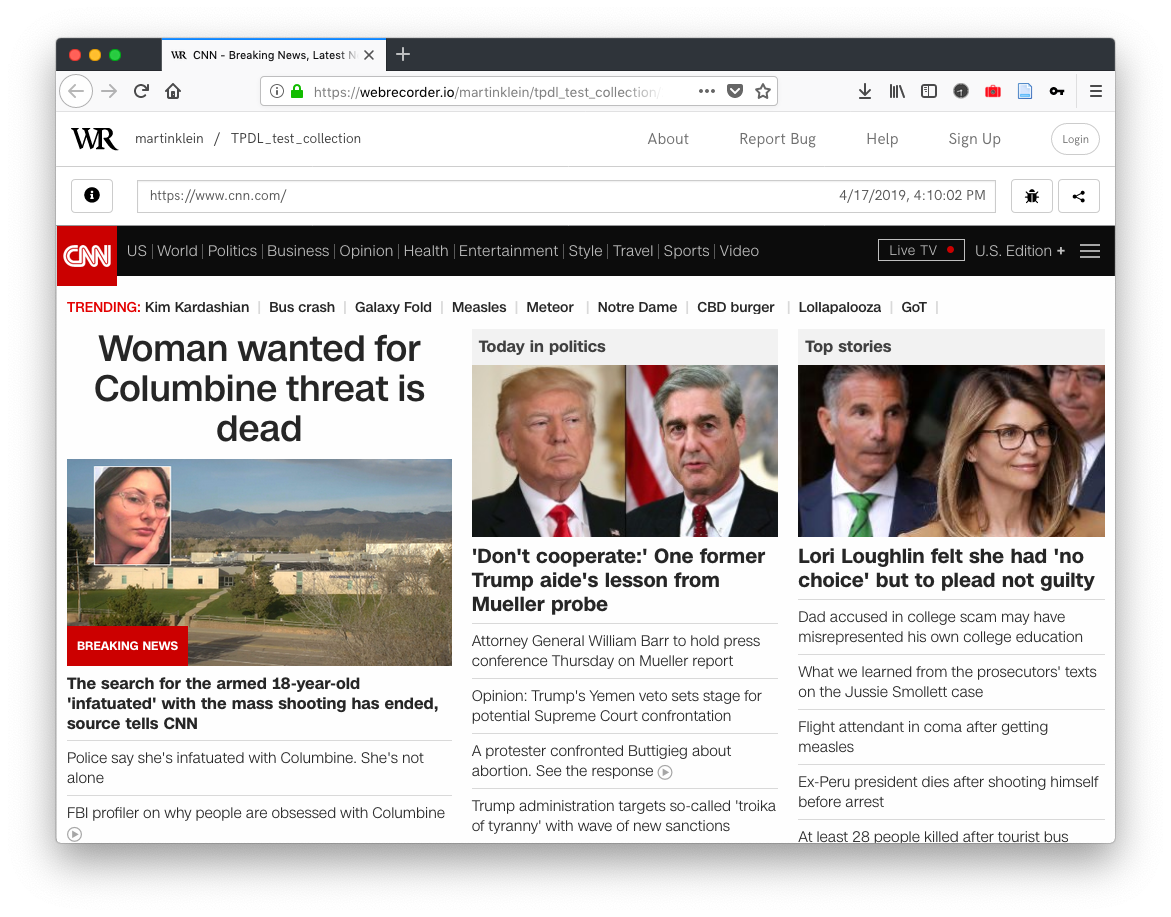}
  \caption{Webrecorder}
  \label{fig:cnn_wr_replay}
 \end{subfigure}
 \caption{Replay of \texttt{cnn.com} Mementos}
 \label{fig:cnn_replay}
\end{figure}

Regarding archival quality, web archiving has become increasingly challenging as a result of the proliferation of dynamic web content that only 
becomes available via activation 
of - typically JavaScript-based - affordances in pages. Web archiving dynamic web content is technically challenging and crawler-driven solutions that 
have been experimented with thus far are resource intensive, slowing down the crawling process \cite{berlin:thesis,brunelle:crawlers}. As such, the 
IA, which focuses on scale, typically does not apply such techniques. The result of this focus on scale over quality is aptly illustrated by the fact 
that the \texttt{cnn.com} website has not been properly archived by the IA since November $2016$ and can not be replayed correctly in the Wayback 
Machine \cite{berlin:cnn}. The screenshot in Figure \ref{fig:cnn_ia_replay} shows the replay of a \texttt{cnn.com} Memento (the archived copy) in the 
IA\footnote{\url{http://web.archive.org/web/20190417195948/https://www.cnn.com/}}.
At the other end of the spectrum, the Webrecorder tool \cite{webrecorder} has emerged, focusing on high-fidelity web archiving. Its value is 
evident in Figure \ref{fig:cnn_wr_replay} showing a screenshot of the replay of a \texttt{cnn.com} Memento created with the
Webrecorder tool\footnote{\url{https://webrecorder.io/martinklein/tpdl_test_collection/20190417221002/https://www.cnn.com/}}.
However, Webrecorder achieves this level of quality via human interaction with the web resource that is to be archived and, as such, can not operate at 
a scale comparable to that of the IA's crawling processes. To date, approaches that can archive at scale and with high-fidelity remain elusive.

In this paper we introduce the Memento Tracer web archiving framework that aims to operate at web scale while also providing high-quality 
web captures. Memento Tracer is a result of the ``Scholarly Orphans'' project, a collaborative effort between the Prototyping Team of the 
Los Alamos National Laboratory Research Library and members of the Web Science and Digital Library group of the Old Dominion University 
Computer Science Department. The project is focused on archiving scholarly artifacts, which are resources that scholars create or deposit 
in productivity portals such as GitHub, Slideshare, or Publons.
Our contributions in this work are two-fold: 
\begin{enumerate}
\item We outline the Memento Tracer concept, detail its architecture, and describe a pilot implementation.
\item We conduct an experimental evaluation of Memento Tracer regarding the scale and quality at which it can archive two resource types that 
present dynamic content challenges.
\end{enumerate}
%
%
Despite the limited scope of the evaluation, we feel that our contributions reveal various attractive characteristics of Memento Tracer, 
which suggest the potential for it to evolve towards a web archiving approach that is able to capture web resources at scale and with high 
quality.
\section{Related Work}
The Internet Archive, in addition to its crawler-based web archiving services offers an archive-on-demand service called ``Save page now''.
This service allows a user to submit a URI to the IA, which will be crawled immediately. Perma.cc and archive.today are two alternatives
that offer very similar services, all of which come with strengths and weaknesses. Perma.cc, for example, requires a user login
to pro-actively create Mementos of submitted URIs and charges a subscription fee that depends on the number of Mementos created per
month. Little is publicly known about archive.today, their technology stack and institutional background but similar to the IA's
``Save page now'' service their capability to handle dynamic web content is limited. The IA has acknowledged this shortcoming and 
introduced a beta version of a more powerful archiving-on-demand service. This service is based on ``brozzler'' \cite{brozzler} and operates 
a Chromium browser to execute dynamic content and therefore, for example, discovers URIs that are generated by JavaScript.
Our first tests did not return reliable results but once the service reaches a more stable state, it should be included in this 
comparative study.

Brunelle et al. \cite{brunelle:crawlers} conducted a study to investigate the balance and implicit overhead between operating a 
``regular'' web crawler such as Heritrix \cite{heritrix} and a headless browser such as PhantomJS \cite{phantomjs} to more reliably 
execute dynamic content. They found that using the more sophisticated crawling approach based on a headless browser resulted in a spike 
of discovered URIs to crawl as well as vastly increased crawl time. 

The Webrecorder tool is made for humans to interact with a web resource and record the interaction into an archival record. Dynamic 
content is typically handled very well and, as long as all essential parts of the resource are interacted with, the archival record 
represents a high-fidelity capture of the live web resource that can be played back, for example, with the Webrecorder 
Player \cite{wr:player}. While archiving with Webrecorder is a manual process, the tool's developers have made some initial steps towards 
automating certain interactions with individual web resources \cite{kreymer:automation}.

Brunelle et al. \cite{brunelle:damage} proposed an automated method to assess the archived quality of web resources. Their algorithm
assigns relative values to embedded resources and depending on the availability of these resources, determines a damage rating. The
authors implemented a web service to assess Memento Damage\footnote{\url{http://memento-damage.cs.odu.edu/}} which we considered
for our quality evaluation but since it does not compare two versions of the same resource but rather analyzes individual resources
separately, it is not applicable for our study.
\section{Memento Tracer Framework}
We introduce a new collaborative approach to capture web publications for archival purposes with the Memento Tracer framework.
The framework is inspired by existing capture approaches yet aims for a new balance between archiving at scale and quality 
of the resulting snapshots. The framework was developed as part of a project that focuses on capturing scholarly artifacts from
productivity portals such as GitHub, Slideshare, Publons, Figshare, Wikipedia, and Stack Overflow. Similar to other existing web 
crawler approaches such as LOCKSS \cite{reich:lockss,rosenthal:lockss}, it uses server-side processes that leverage the insight that 
web publications in a given portal are typically based on the same template and share features such as layout and interactive affordances. 

Similar to the Webrecorder tool, a human helps achieve high quality captures and determines the boundary of the to be archived
resource. However, with Memento Tracer, heuristics that apply to an entire class of web publications are recorded, not individual
web publications. These heuristics can collaboratively be created by curators and deposited in a shared community repository. 
When the server-side capture processes come across a web publication of a class for which heuristics are available, they can be applied, 
yielding captures that are aligned with the curators' instructions.
\subsection{Framework} \label{sec:framework}
\begin{figure}[t]
 \includegraphics[width=\textwidth]{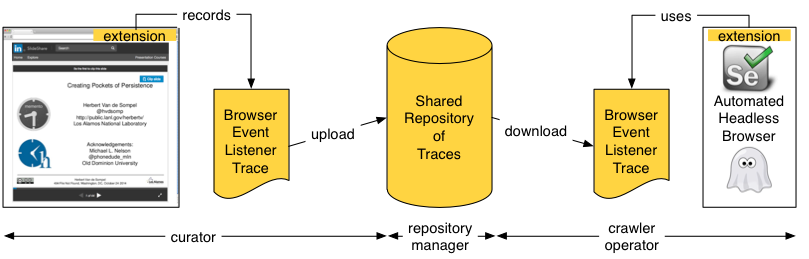}
 \caption{Memento Tracer framework} 
 \label{fig:memento_tracer}
\end{figure}
Figure \ref{fig:memento_tracer} visualizes the Memento Tracer concept and its three main components from left to right.
Below we describe the framework in detail using the task of archiving slide decks from Slideshare as an example.

\textbf{A Browser Extension} 
The first step in the framework begins when a curator navigates to a web page representative for a class of resources in that 
portal, for example the landing page of a Slideshare presentation, and activates the browser extension. By interacting with the web 
page (clicking through the slides, downloading the entire slide deck, etc), the curator creates a \textit{trace} that, in an abstract 
manner, describes the artifact to be archived. The extension does not record actual resources or URLs that are traversed by the curator. 
Rather it captures interactions in terms that uniquely identify the page's elements that are being interacted with, for example by means 
of their class ID or XPath. 
The extension's recording of a page's elements is inspired by the Selenium IDE\footnote{\url{https://www.seleniumhq.org/selenium-ide/}}, 
which is an open source record and playback test automation suite for the web. 
Since all pages of the same class in the same portal are typically based on the same template, the resulting traces apply across all pages 
of the class rather than to a specific page only. In our example the created trace is valid for all slide decks in Slideshare. 
Currently, the extension is able to record simple mouse-clicks, clicks on all links in a certain user interface component, and repeated 
clicks. The latter is especially useful when navigating through all slides of a presentation or paginating through multi-page blog posts 
or manuals.
%
%
The created trace also indicates the URL pattern to which the trace applies and provenance information including the resource on which the 
trace was created and the user agent used to create it. When the layout or affordances for a particular class of web publications change, 
a new trace needs to be recorded to ensure it is valid for all changed publications.
In contrast to crawler-based approaches but similar to the Webrecorder concept, with Memento Tracer a curator is in charge of determining the
desired components of a web resource that is to be archived.
The fact that a trace can automatically be applied to all artifacts of the same class represents a major scalability advantage over other 
human-driven approaches such as Webrecorder. For each resource, even in the same class and portal, Webrecorder requires all interactions to 
be executed.

\textbf{A Shared Repository}
After a curator has successfully recorded a trace, she can share it with the community via a publicly accessible repository, thereby crowdsourcing 
the web curator task. The shared repository allows for reuse and refinement of existing traces. Hence, anyone in 
the community can utilize a trace created by another curator, for example the aforementioned Slideshare trace, to capture other slide decks.
Since the perspective of what the essence of a web publication is may differ from one curator to the next \cite{poursardar:perception}, the 
repository supports multiple traces for a specific class of pages. Each can be unambiguously identified in the repository. In addition, since the 
layout of pages evolves over time, traces will need updating, making version support by the repository essential. 
Given these requirements, we consider GitHub a suitable host for the shared repository.
Traces available to the community for reuse and refinement in addition to versioning in a shared repository further increases the scalability 
of the Memento Tracer approach. 

\textbf{A Headless Browser Application} 
To generate web captures, the Memento Tracer framework assumes a setup consisting of 
a WebDriver\footnote{Selenium WebDriver: \url{https://www.seleniumhq.org/}},
a headless browser\footnote{Headless Chrome: \url{https://chromium.googlesource.com/chromium/src/+/lkgr/headless/README.md}}, and
a capturing tool\footnote{WarcProxy: \url{https://github.com/internetarchive/warcprox}}. 
We developed a parser for the WebDriver (based on the Selenium WebDriver's API) that translates the content of a trace into instructions 
(JavaScript code) for the headless browser to emulate the interactions with the web resource as captured by the browser extension in the trace. 
The capturing tool writes resources navigated by the headless browser to WARC files \cite{iso:warc}. 
When this fully automated capture setup comes across a web resource of a class for which a trace is available, the trace will be invoked to guide 
the capturing of the resource. 
This functionality of capturing resources based on traces guarantees high-fidelity archived resources, which is a major advantage over, for example, 
the IA's automated crawling approaches. 
\subsection{Pilot} \label{sec:pilot}
We used a pilot implementation of the above described framework to capture artifacts deposited by $16$ researchers in $11$ productivity portals. 
We created a trace for a sample resource of each of the portals and used those traces to guide the capturing process of artifacts deposited in 
the respective portals by the researchers. The application at \url{https://myresearch.institute} provides an overview of artifacts captured since 
August $2018$. The application offers different views on the collection of captured artifacts, such as by capture date, by researcher, and by 
productivity portal. A landing page per captured 
artifact\footnote{For example: \url{https://myresearch.institute/event/e7e8fcc4e8c14392af1c264295d6268a/}} provides basic metadata about the 
artifact as well as links to the WARC file resulting from the capture and to a replay of the captured artifact. More information about the capture 
pipeline in which Memento Tracer was used is available via the \texttt{About} page\footnote{\url{https://myresearch.institute/about/}}. 
\section{Experiment Design} \label{sec:experiment_design}
We evaluate our Memento Tracer framework in two dimensions: archival quality and scalability. To assess archival quality, we compare its 
performance against Mementos created with Webrecorder, the tool designed to create high-fidelity captures. To evaluate scalability, we conduct 
two experiments. The first assesses the extent to which Memento Tracer can generate quality captures for a large set of web resources. The second 
compares the time Memento Tracer and an automated crawling framework designed for scale require to create captures.

Quality of archived web resources is not trivial to measure. Different replay systems of Mementos may vary in performance
and individuals' perception of what the essence of a resource is and hence which part of the resource needs to be part of the archived
record may differ \cite{poursardar:perception}. Rather than trying to find a compromise between these arguably subjective aspects, we 
decided to focus our quality assessment on the extent to which URIs that should be captured according to curatorial decisions are actually 
captured.
In order to create a baseline of the number of URIs we expect in a Memento, we analyze the live web version of each resource. 
We expect a high-quality archival record to contain at least the same number of URIs as its live web version. We are aware that this 
comparison may not capture all dimensions of quality. For example, a CSS that is missing from a captured resource may have a more detrimental 
impact on the ``look and feel'' of the replay than a missing image. However, this process enables us to automatically compare a dataset of live 
web resources with their corresponding Mementos.
\subsection{Data Gathering} \label{ssec:data_gathering}
Memento Tracer is a result of our Scholarly Orphans project, where our focus is on archiving scholarly artifacts that researchers deposit in 
web productivity portals. We generated a dataset that consists of resources from two such portals selected because they present interesting 
web archiving challenges to analyze the performance of our novel web archiving framework. This dataset is applicable to investigate web archiving 
quality and scale. 
The first portal from which we obtained resources is GitHub. 
Its API does not offer the functionality to randomly select Github resources. We therefore decided to utilize the news and podcast 
platform \texttt{\url{https://changelog.com/}} and its digest of GitHub repositories published daily since January 1st $2015$.
Changelog's digest consists of the most popular GitHub repositories on a given day as measured by the number of stars received. These
repositories are further distinguished between popular overall, popular overall but making the list for the first time, and popular
overall but newly created repositories. We focused on the latter two categories as this ensures we obtain established repositories, while 
also decreasing the chances of obtaining duplicates. Furthermore, newly created repositories are included.
This mixture of GitHub repositories, while somewhat biased towards popularity (given the number of received stars or ``likes''), serves 
as our sample set of resources. In total, we obtained $17,646$ URIs of GitHub repositories. In order to conduct an accurate analysis of
archiving quality, we need to ensure that our comparisons are based on the live web versions that were used to create Mementos, which,
in fact, may no longer be the same version by the time we conduct our comparisons. We therefore use the GitHub API to identify the 
time-specific last commit URI of each repository and use these URIs, for which the repository content is fixed.

Our second dataset consists of resources from Slideshare. In order to obtain a random sample, we use the portal's Explore 
feature\footnote{\url{https://www.slideshare.net/explore}} to obtain a sample of slide decks. 
Given this source, our dataset is clearly also biased towards popularity and Slideshare's selection algorithm. However, the algorithm to 
select slide decks and feature them on the Explore site is entirely opaque to us. In addition, this process guarantees a broad variety
of subjects under which the slide decks are classified, making our sampling results an applicable dataset.
Since Slideshare creates a new URI for each uploaded and updated slide deck, there is no concern that the resource on the live web will 
change throughout our experiment. In total, we obtain $12,280$ URIs of distinct slide decks for this dataset.
\subsection{Traces for Dataset Resources}
We use the Memento Tracer Chrome extension to create a trace for GitHub repositories as well as for Slideshare slide decks.
For this step we mimic the decision making of a curator and determine which parts of the resources are essential to be captured 
and archived. According to these curatorial decisions, the trace created for GitHub repositories includes all files and top level directories 
listed in the repository as well as the downloadable ZIP file containing the entire repository\footnote{A screencast of the Memento Tracer 
Chrome extension and the interactions with a GitHub repository recorded into a trace is available 
at: \url{https://doi.org/10.6084/m9.figshare.8049839.v1}}. 
Guided by this trace\footnote{The trace is available at: \url{https://doi.org/10.6084/m9.figshare.8024612}}, all GitHub repositories 
archived with Memento Tracer should therefore, when replayed, contain all repository files as well as the ZIP file.
The trace created for Slideshare guides the capturing process to include all slides as well as all notes per slide 
deck\footnote{The trace is available at: \url{https://doi.org/10.6084/m9.figshare.8024615}}.

These curatorial decisions allow us to precisely and automatically determine the number of URIs of interest contained in the 
live web version of each resource. For this purpose, our evaluation program loads the live web resource in a browser and interacts 
with it to count the number of URIs expected according to the curatorial decisions made. With this process we determine that 
each file and top-level directory in a GitHub repository as well as the Zip file has a distinct URI. Similarly, each slide in a 
Slideshare slide deck as well as its associated note has a unique URI. 
%
%
%
%
%
\section{Experiments and Results}
\subsection{Archival Quality}
With our baseline of live web URIs in place, we can compare Mementos created with different archiving frameworks.
We use the same evaluation program we used to assess the number of URIs from live web resources to assess the number of URIs 
in Mementos. To conduct this comparison, we create the following subsets derived from our dataset introduced in 
Section \ref{ssec:data_gathering}.
We randomly pick $100$ GitHub repositories and $100$ Slideshare slide decks and use the Webrecorder tool to manually create respective 
Mementos. Being very familiar with the Webrecorder tool, we applied the same curatorial criteria that we used to record traces for the 
two productivity portals. For GitHub repositories, this means we click on every single file in a GitHub repository in order to capture 
these resources and the ``Clone or Download'' button in order to capture the ZIP file of the repository. For slide decks this means we 
click on the ``Next'' button as many times as necessary to capture all slides in the deck and on each of the included notes. 
Since this manual process is rather time consuming, we limit the size of this dataset to $100$ repositories and $100$ slide decks.
In addition, with a trace recorded for GitHub repositories and for slide decks on Slideshare, we use the Memento Tracer framework 
to capture the same $100$ GitHub repositories and $100$ Slideshare slide decks.
\begin{figure}[t!]
 \centering
 \begin{subfigure}[h]{0.49\textwidth}
  \includegraphics[scale=0.37]{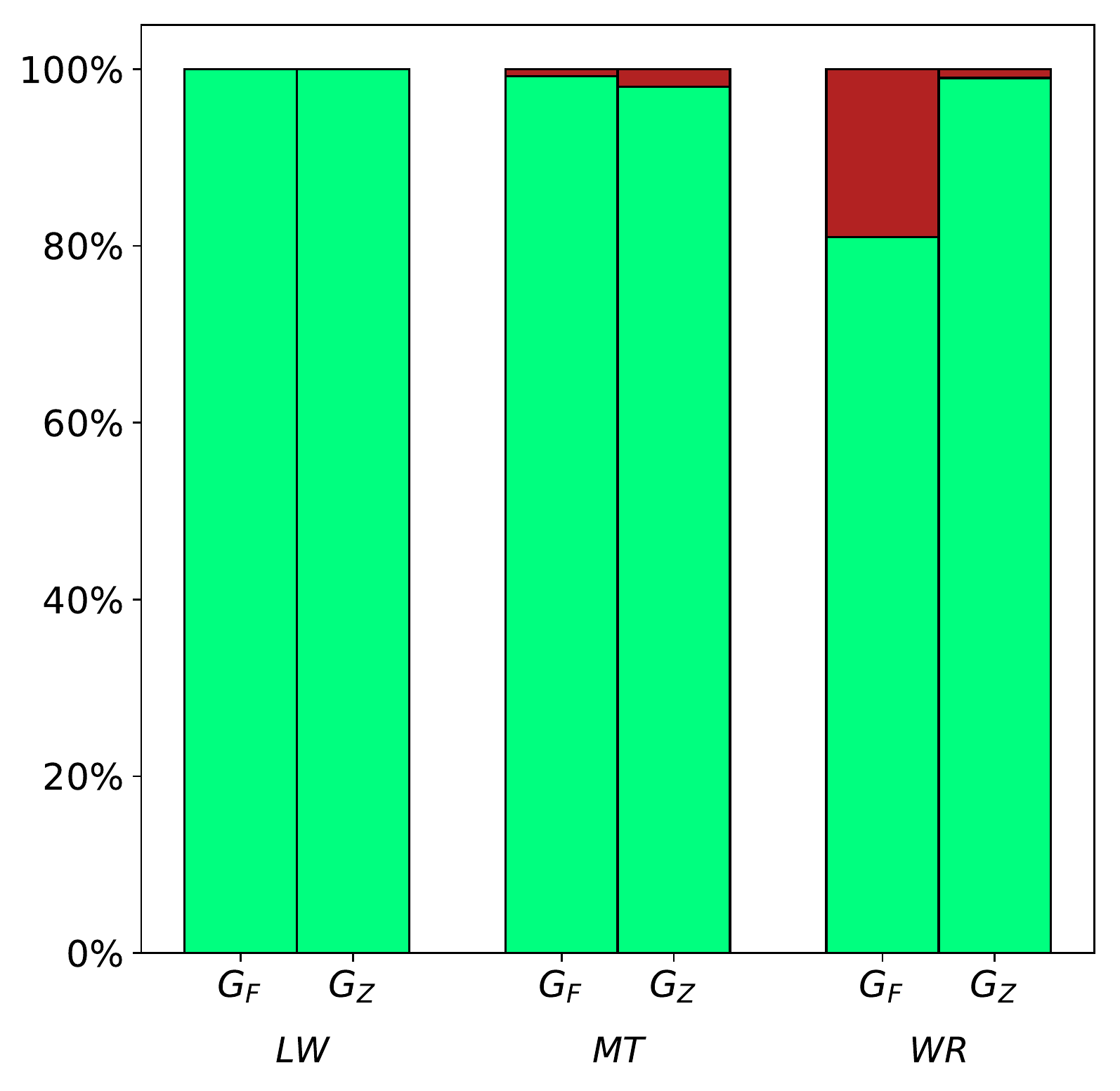}
  \caption{GitHub URIs, each left bar represents file URIs, the right ZIP file URIs} 
  \label{fig:100_comparison_gh}
 \end{subfigure}
 \begin{subfigure}[h]{0.49\textwidth}
 \includegraphics[scale=0.37]{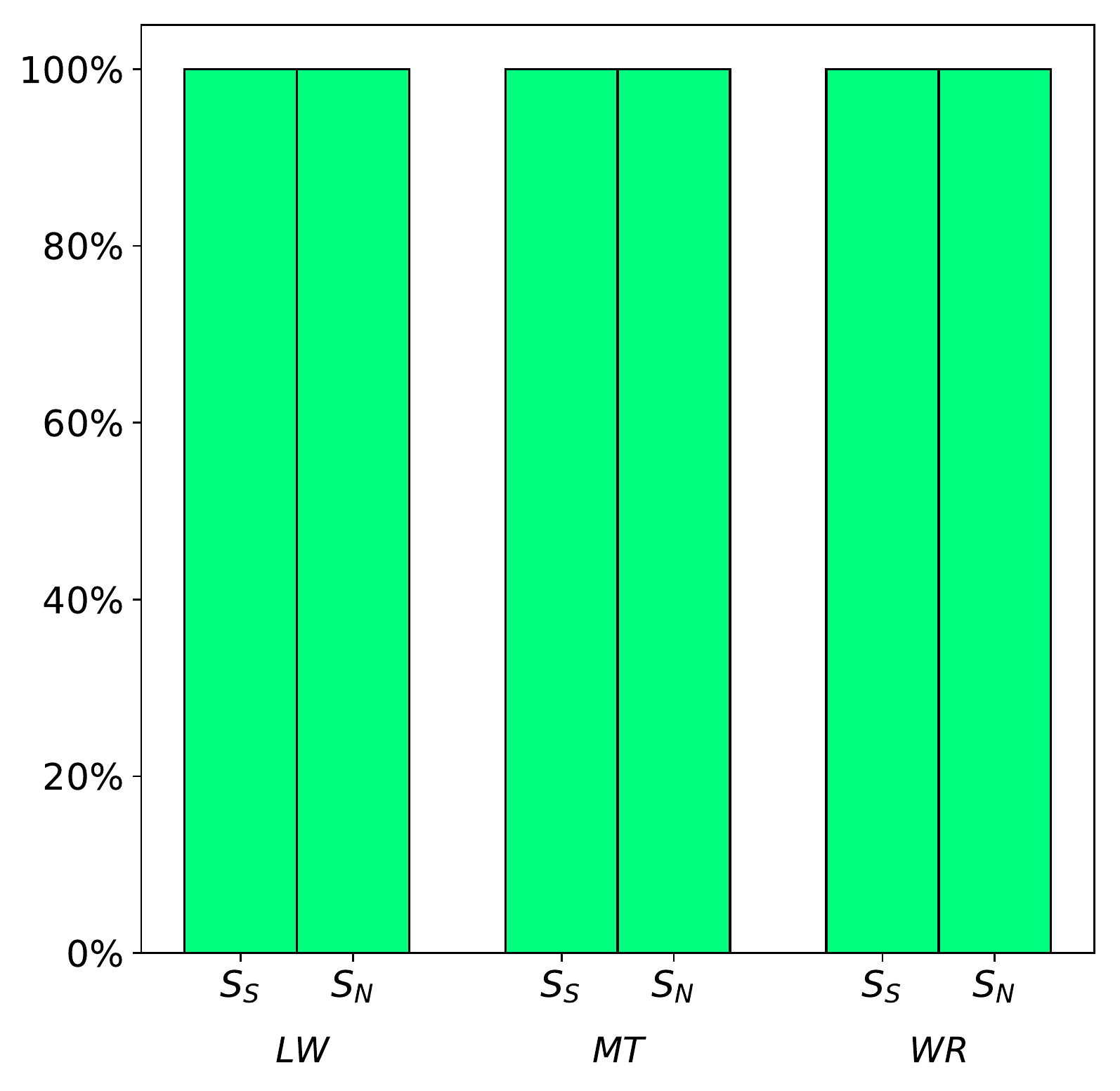}
  \caption{Slideshare URIs, each left bar represents slide URIs, the right notes URIs} 
  \label{fig:100_comparison_ss}
 \end{subfigure}
 \caption{Relative number of URIs from live web, Memento Tracer, and Webrecorder Mementos.
Green represents available, red unavailable resources.}
 \label{fig:100_comparison}
\end{figure}

Our first analysis is based on whether all expected URIs are contained in the archived record. We expect all URIs for files in GitHub 
repositories (each file has a distinct URI) and one URI for the repository ZIP file in addition to the URI of the repository itself. 
For Slideshare we expect all URIs for slides (each slide has a distinct URI) and all URIs for notes (each note has a distinct URI) plus 
the URI of the slide deck page itself.
Since the URIs of the repositories and slide deck pages are all available on the live web and in all Mementos, we exclude them going
forward and only focus on URIs of the component resources that are of interest according to our curatorial decisions.

Figure \ref{fig:100_comparison} displays the results of the analysis based on the total of $200$ URIs sampled from the overall dataset. 
The relative numbers of URIs are represented on the y-axis and the corresponding sources (LW, MT, WR) are shown on the x-axis.
The size of the green portion of a bar indicates the number of URIs available and the red portion shows unavailable URIs.
Figure \ref{fig:100_comparison_gh} displays the GitHub URIs where $G_F$ corresponds to repository file URIs and $G_Z$ to ZIP file 
URIs. Figure \ref{fig:100_comparison_ss} shows the Slideshare URIs with $S_S$ representing URIs of slides and $S_N$ URIs of notes.
We can immediately make a few observations from these graphs. As expected, the ratio of URIs in Webrecorder Mementos is very similar 
to the live web versions. Generally, more than $95\%$ of URIs are available, which confirms the tool's reputation of delivering 
high-fidelity captures. 
We also notice very high ratios of available URIs for Mementos created with Memento Tracer. In fact, the ratio of available URIs 
in Memento Tracer Mementos is at times even higher than the ratio in Webrecorder Mementos.
The drop in available URIs from Webrecorder GitHub repository file URIs can potentially be explained by observed network errors 
while creating the archival snapshots as well as possible human errors, for example forgetting to click on a file. Both points favor 
an automated framework to capture web resources. Such a process can detect network errors, try the capture again, and is not subject
to human errors.

These findings strongly support our claim that Memento Tracer captures are of high quality, they are
comparable to if not better than Webrecorder Mementos, and very closely align with their live web versions.
\begin{figure}[t!]
 \centering
 \begin{subfigure}[h]{0.49\textwidth}
  \includegraphics[scale=0.37]{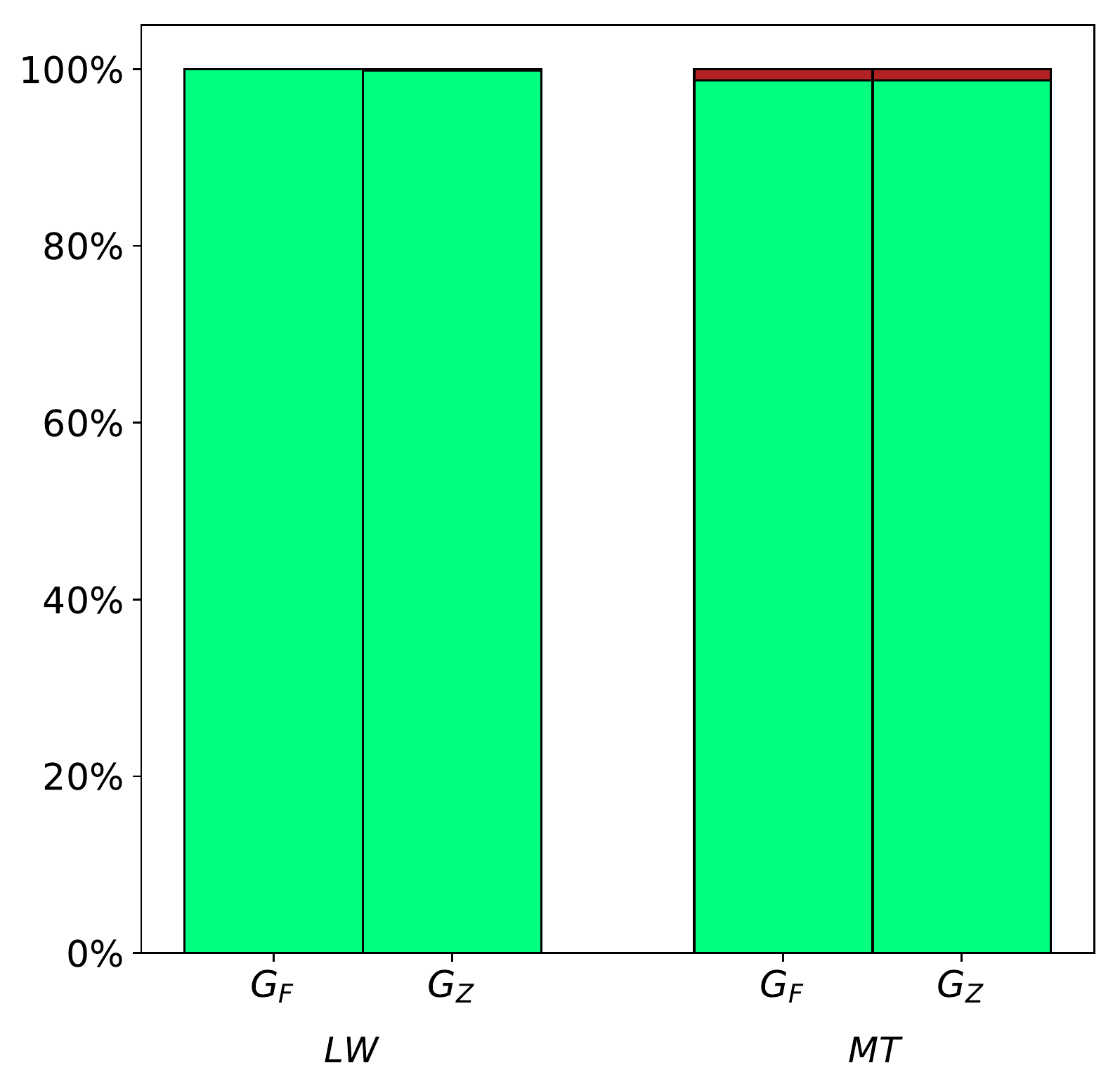}
  \caption{GitHub URIs, files left, ZIP right}
  \label{fig:all_comparison_gh}
 \end{subfigure}
 \begin{subfigure}[h]{0.49\textwidth}
 \includegraphics[scale=0.37]{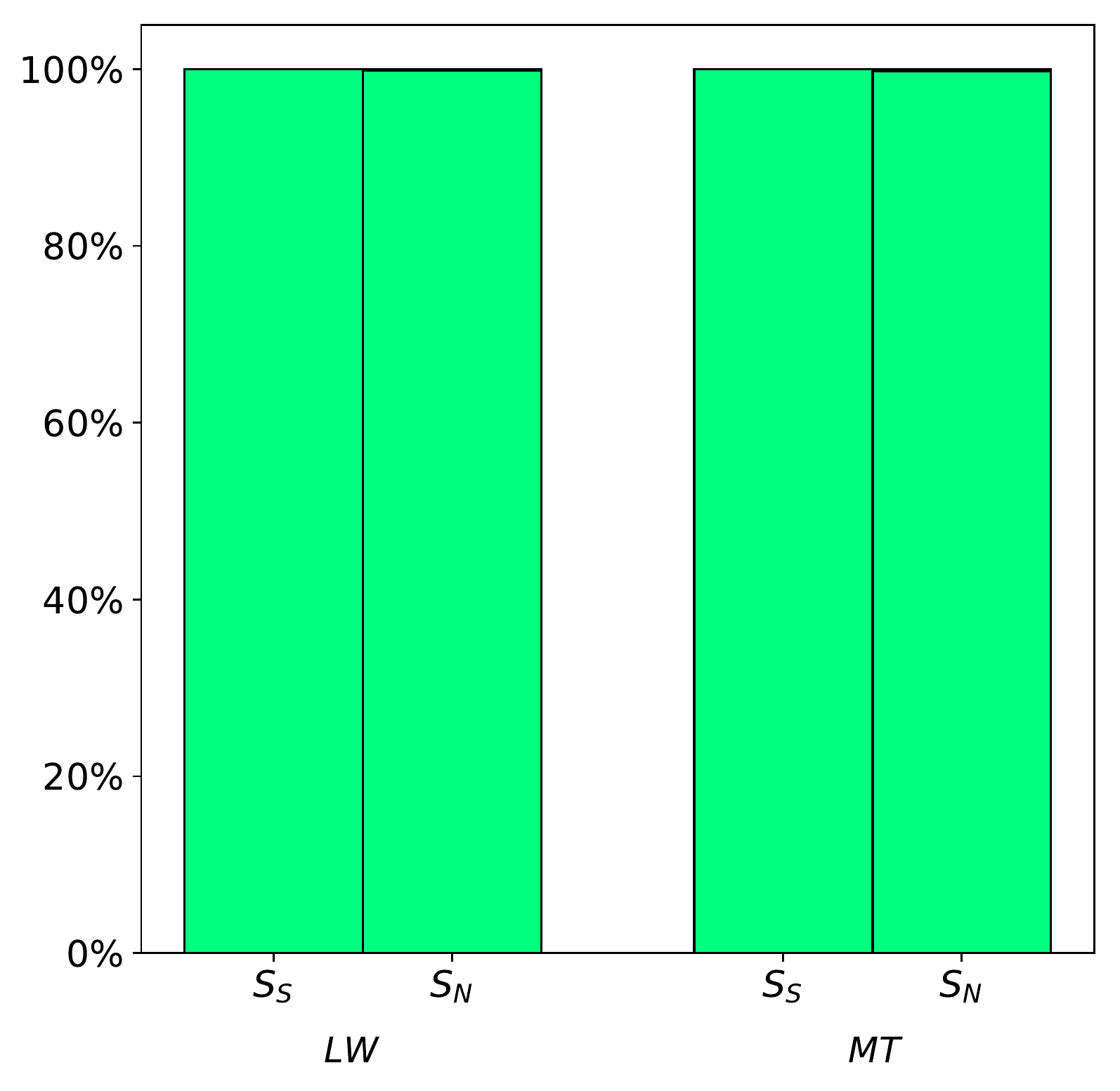}
  \caption{Slideshare URIs, slides left, notes right} 
  \label{fig:all_comparison_ss}
 \end{subfigure}
 \caption{Relative number of URIs from live web and Memento Tracer Mementos.
Green represents available, red unavailable resources.}
 \label{fig:all_comparison}
\end{figure}
\subsection{Quality at Scale}
Using our subset in previous experiments, we established that Memento Tracer Mementos are of very high quality, when compared 
to their live web versions (and Webrecorder Mementos). We are now interested in analyzing whether Memento Tracer keeps delivering 
quality when operating at scale. We use the framework to capture all resources in our dataset; $17,646$ Memento Tracer Mementos of 
GitHub repositories as well as all $12,280$ Memento Tracer Mementos of Slideshare slide decks. This translates to a dataset increase 
of more than two orders of magnitude ($100$ vs. $17,646$ repositories and $100$ vs. $12,280$ slide decks). If we find a high level 
of similarity in terms of available URIs observed in the live web versions and Memento Tracer Mementos, we can confidently state that, 
even at scale, the Memento Tracer approach provides high-quality captures.

Figure \ref{fig:all_comparison} shows, in concept similar to Figure \ref{fig:100_comparison}, the results of this large-scale analysis.
Figure \ref{fig:all_comparison_gh} represents the GitHub URI ratios and Figure \ref{fig:all_comparison_ss} the Slideshare ratios.
We can barely see a red portion in any of the bars, indicating that the same almost $100\%$ of URIs available in live web versions 
are also available in the corresponding Memento Tracer Mementos. 

Table \ref{tab:lw_mt_similarity} provides insight into the comparison between live web versions and Memento Tracer Mementos from the 
granularity level of GitHub repositories and Slideshare slide decks. The second row shows the percentage of Memento Tracer Memento GitHub 
repositories with $x$ percent (top row) of available URIs from the corresponding repositories' live web version. For example, we can see 
that $0.64$ of Memento Tracer Memento repositories contain zero URIs from their live web version and $92.83\%$ contain all $100\%$ of 
available URIs. The third row shows the same data if we only consider repository file URIs and the fourth when only considering ZIP file 
URIs. Since there is only one ZIP file per repository, this data is binary. We can observe that Memento Tracer does very well overall, and 
slightly better for ZIP files ($98.7\%$ vs. $93.29\%$). 
The most likely reasons are that temporary network issues not caught by the automatic capture process prevented the Memento Tracer framework 
from archiving all file URIs.
Rows five through seven show the same data for Slideshare slide decks. We find that Memento Tracer does even better there and almost perfect 
($99.9\%$) for slides. The percentage for notes is also very high, at $98.58\%$.
\begin{table}[t!]
 \centering
 \caption{Percentage of GitHub repositories with $x$ percent of available URIs from the corresponding repositories' live web version}
 \label{tab:lw_mt_similarity}
 \begin{tabular}{|l|l||c|c|c|c|c|c|c|c|c|c|c|} \hline
& $x$ & $0$ & $10$ & $20$ & $30$ & $40$ & $50$ & $60$ & $70$ & $80$ & $90$ & $\textbf{100}$ \\ \hline \hline
\multirow{3}{*}{GitHub} & All & $0.64$ & $99.36$ & $97.18$ & $96.92$ & $96.86$ & $96.83$ & $96.7$ & $96.36$ & $95.91$ & $94.81$ & $\textbf{92.83}$ \\ 
\cline{2-13}
& Files & $2.99$ & $97.01$ & $97.0$ & $96.97$ & $96.94$ & $96.91$ & $96.75$ & $96.43$ & $95.97$ & $94.95$ & $\textbf{93.29}$ \\
\cline{2-13}
& ZIP & $1.3$ & n/a & n/a & n/a & n/a & n/a & n/a & n/a & n/a & n/a & $\textbf{98.7}$ \\ \hline \hline
\multirow{3}{*}{Slideshare} & All & $0.15$ & $99.85$ & $99.85$ & $99.85$ & $99.85$ & $99.83$ & $99.71$ & $99.71$ & $99.7$ & $99.65$ & $\textbf{98.67}$ \\ 
\cline{2-13}
& Slides & $0.1$ & $99.9$ & $99.9$ & $99.9$ & $99.9$ & $99.9$ & $99.9$ & $99.9$ & $99.9$ & $99.9$ & $\textbf{99.9}$ \\
\cline{2-13}
& Notes & $0.28$ & $99.72$ & $99.71$ & $99.71$ & $99.71$ & $99.71$ & $99.68$ & $99.65$ & $99.62$ & $99.56$ & $\textbf{98.58}$ \\ \hline
 \end{tabular}
\end{table}

These findings exceed our expectation and confirm that the Memento Tracer framework, even at large scale, archives web 
resources with high-quality.
\subsection{Memento Tracer Overhead}
\begin{figure}[t!]
 \centering
 \includegraphics[width=\textwidth]{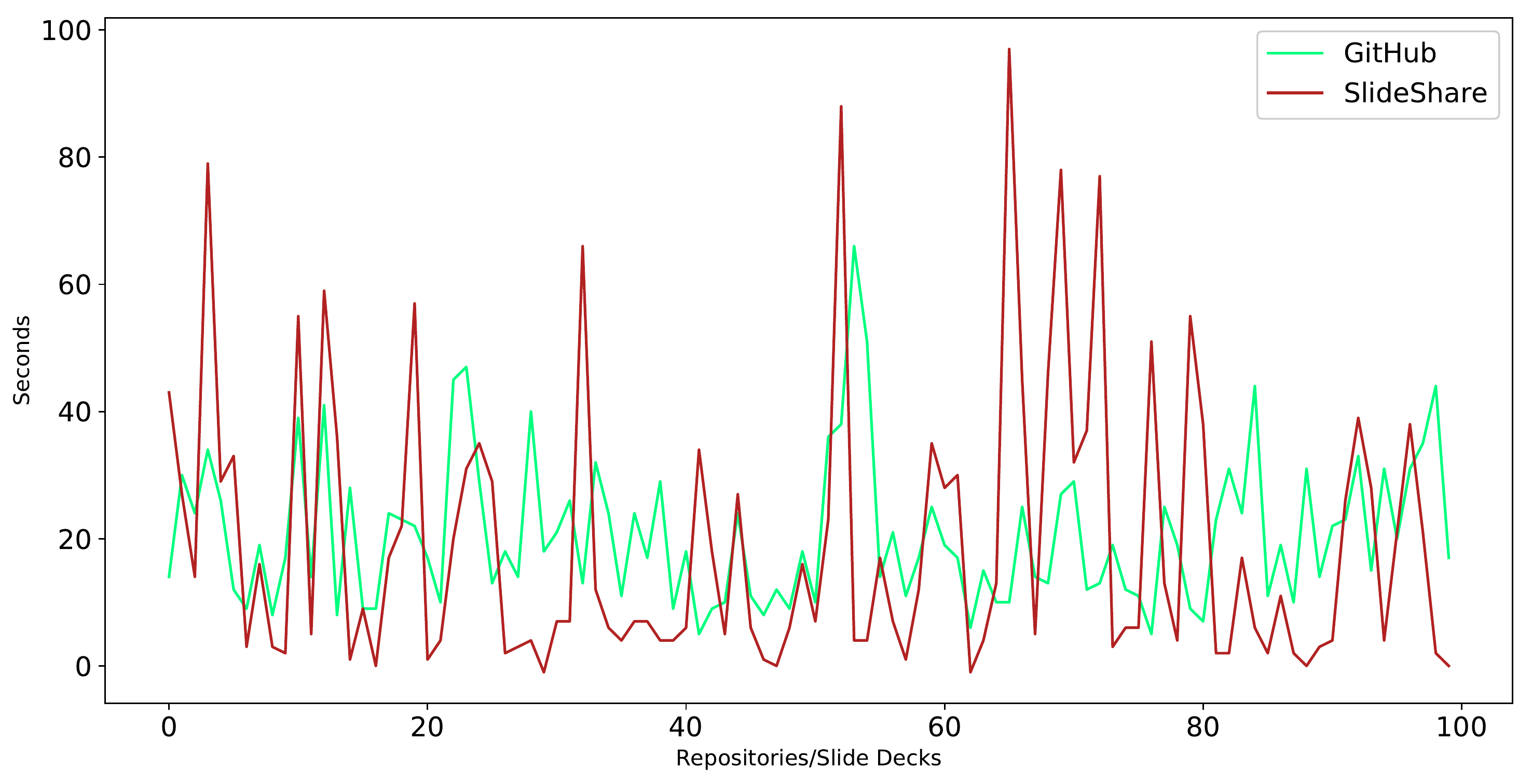}
 \caption{Time deltas between Memento Tracer and a simple web crawler} 
 \label{fig:time_delta}
\end{figure}
We are further interested in the overhead of the WebDriver- and headless browser-based capture approach in the Memento 
Tracer framework. Simply comparing runtimes of Memento Tracer versus another automatic web archiving framework such as Heritrix 
would not be fair as a crawler would not discover the same URIs since it can not cope with some dynamic affordances.
Instead, we extract all URIs captured by the Memento Tracer framework while creating Mementos of the initial subset of $100$ GitHub 
URIs and $100$ Slideshare URIs. This amounts to a total of $29,205$ extracted URIs from the GitHub subset and $61,346$ URIs from 
SlideShare and we crawl these URIs with a simple Python-based crawler. 
Our simple crawler builds on the popular Python \texttt{Requests} library to perform HTTP GET requests against the URIs and is 
configured to resemble a Chrome browser (specified user agent, set timeout values, etc). We also used $16$ threads to parallelize 
these HTTP requests, speed up the crawling process, and emulate a production crawling framework operating at scale.
The advantage of this process is that the crawler simply captures URIs and its runtime therefore provides the minimal time needed to 
capture the same resources as Memento Tracer.
We compare both runtimes and present the delta in Figure \ref{fig:time_delta}. All deltas (per GitHub and Slideshare URI)
are positive (y-axis), which means Memento Tracer in all instances takes longer than the simple crawler. This finding is not 
surprising given Memento Tracer's overhead of running and controlling a browser for each URI. We see quite a variance from around 
$5$ seconds to just under $100$ seconds for Slideshare URIs. On average, based on our subset, GitHub URIs take $19.74$ seconds longer to 
be captured with Memento Tracer and Slideshare URIs take $20.75$ seconds longer. If we extrapolate this average to our entire dataset, 
the GitHub portion would take $97$ hours longer and Slideshare $71$ hours. 

While this may sound like a lot, we highlight two arguments for why these numbers are reasonable; First, on average, the Memento Tracer 
framework is $10.17$ times slower than the simple crawler. In contrast, Brunelle et al. \cite{brunelle:crawlers} found their headless 
browser approach to be $38.9$ times slower than a common crawler, so we see a significant decrease in crawl time. Second, simple 
automatic crawlers would not even discover a lot of the captured URIs, so speed is not the only factor.
Runtimes can vary, depend on network speeds, and potential framework and crawler optimizations but objectively, these numbers provide 
insight into the cost (extra time) involved in automatic high-quality archiving. 
\section{Discussion and Future Work}
Memento Tracer was developed as part of the Scholarly Orphans project, which focuses on artifacts that researchers deposit in a 
limited set of web productivity portals. Investigating the framework's applicability and merit beyond that scope remains for future 
work. We anticipate limitations imposed by the Chrome extension to create a trace for resources and limited value of the automated 
approach for web resources not based on common templates. 

Traces created with our browser extension are currently expressed in a non-standardized manner. In order to enable interoperability 
between traces and other capture frameworks such as Puppeteer\footnote{\url{https://github.com/GoogleChrome/puppeteer}}, a standard
language needs to be devised to express interactions.

We are exploring alternate framework components to further stabilize our pilot implementation of the framework. Both the headless browser
and the WarcProxy tool have proven unreliable at times.
\section{Conclusion}
In this paper we introduced Memento Tracer - a framework that provides high quality captures of web resources.
Memento Tracer puts the curator in charge of determining the desired components of a to be archived web resource and takes 
advantage of frequently reused patterns in online productivity portals. We conducted experiments that show that Memento Tracer 
delivers high archival quality and that it can even outperform the Webrecorder tool that was designed for high 
fidelity captures. We have further shown that Memento Tracer captures web resources at high quality, even when operated 
at scale. The technical complexity of the framework, however, comes at a cost. In our experimental setup, compared to a simple 
crawling framework, Memento Tracer takes around $20$ seconds longer to capture a single URI. 
Our findings prove the feasibility and highlight the potential of the Memento Tracer approach. As such, the contributions of this work
should be considered a next step towards balancing quality and scalability for web archiving.
\section{Acknowledgement}
This work is supported in part by The Andrew W. Mellon Foundation grant 11600663.
%
%
%
%
%

\begin{thebibliography}{10}
\providecommand{\url}[1]{\texttt{#1}}
\providecommand{\urlprefix}{URL }
\providecommand{\doi}[1]{https://doi.org/#1}

\bibitem{iso:warc}
{ISO} 28500:2017 - information and documentation -- {WARC} file format,
  \url{https://www.iso.org/standard/68004.html}

\bibitem{ukna}
Archives, U.N.: {The National Archives},
  \url{https://www.nationalarchives.gov.uk/}

\bibitem{berlin:cnn}
Berlin, J.: {CNN.com Has Been Unarchivable Since November 1st, 2016},
  \url{https://ws-dl.blogspot.com/2017/01/2017-01-20-cnncom-has-been-unarchivable.html}

\bibitem{berlin:thesis}
Berlin, J.A.: To relive the web: A framework for the transformation and
  archival replay of web pages (2018), master of Science (MS), Thesis, Computer
  Science, Old Dominion University

\bibitem{brunelle:crawlers}
{Brunelle}, J.F., {Weigle}, M.C., {Nelson}, M.L.: Archival crawlers and
  javascript: Discover more stuff but crawl more slowly. In: 2017 ACM/IEEE
  Joint Conference on Digital Libraries (JCDL). pp. 1--10 (2017)

\bibitem{brunelle:damage}
Brunelle, J.F., Kelly, M., SalahEldeen, H., Weigle, M.C., Nelson, M.L.: Not all
  mementos are created equal: Measuring the impact of missing resources. In:
  Proceedings of the 14th ACM/IEEE-CS Joint Conference on Digital Libraries.
  pp. 321--330 (2014)

\bibitem{phantomjs}
Hidayat, A.: {PhantomJS}, \url{https://github.com/ariya/phantomjs}

\bibitem{brozzler}
{Internet Archive}: Brozzler, \url{https://github.com/internetarchive/brozzler}

\bibitem{heritrix}
{Internet Archive}: Heritrix web crawler,
  \url{https://github.com/internetarchive/heritrix3}

\bibitem{wayback}
{Internet Archive}: Wayback machine, \url{http://web.archive.org/}

\bibitem{kahle:tweet}
Kahle, B.: Wayback rising!,
  \url{https://twitter.com/brewster_kahle/status/1118172506777509890}

\bibitem{kreymer:automation}
Kreymer, I.: A prototype of automated web archiving, emulation and server
  preservation,
  \url{https://blog.webrecorder.io/2018/08/28/automation-emulation-server-preserve.html}

\bibitem{webrecorder}
Kreymer, I.: Webrecorder, \url{https://github.com/webrecorder/webrecorder}

\bibitem{wr:player}
Kreymer, I.: Webrecorder player,
  \url{https://github.com/webrecorder/webrecorder-player}

\bibitem{nla}
{National Library of Australia}: Trove, \url{https://trove.nla.gov.au/}

\bibitem{poursardar:perception}
{Poursardar}, F., {Shipman}, F.: How perceptions of web resource boundaries
  differ for institutional and personal archives. In: 2018 IEEE International
  Conference on Information Reuse and Integration (IRI). pp. 126--129 (2018)

\bibitem{reich:lockss}
Reich, V., Rosenthal, D.S.H.: {LOCKSS}: A permanent web publishing and access
  system. D-Lib Magazine  \textbf{7}(6) (2001)

\bibitem{rosenthal:lockss}
Rosenthal, D.S.H., Vargas, D.L., Lipkis, T.A., Griffin, C.T.: {Enhancing the
  LOCKSS Digital Preservation Technology}. D-Lib Magazine  \textbf{21}(9/10)
  (2015). \doi{10.1045/september2015-rosenthal}

\end{thebibliography}
%
%

%
%
\end{document}